\documentclass{elsart}

\usepackage{amssymb}
\usepackage{amsmath}
\usepackage{epsfig}
\usepackage{amssymb}
\usepackage{epsf}
\usepackage{epsfig}
\usepackage{color}
\usepackage{ifpdf}
\usepackage{cite}
\usepackage{cases}
\usepackage{subfigure}
\usepackage{multirow}

\newcommand{\ba}{\begin{eqnarray}}
\newcommand{\ea}{\end{eqnarray}}
\newcommand{\be}{\begin{equation}}
\newcommand{\ee}{\end{equation}}
\newcommand{\bd}{\begin{displaymath}}
\newcommand{\ed}{\end{displaymath}}
\newcommand{\een}{\nonumber\end{equation}}
\newcommand{\bea}{\begin{eqnarray}}
\newcommand{\eean}{\nonumber\end{eqnarray}}
\newcommand{\eea}{\end{eqnarray}}

\newcommand{\hf}{\frac{1}{2}}

\def\eq#1{Eq.~(\ref{#1})}
\def\eqs#1{Eqs.~(\ref{#1})}
\def\fig#1{Fig.~\ref{#1}}

\def\tab#1{Table~\ref{#1}}

\newcommand{\gap}{\hspace{10pt}}

\newcommand{\mev}{\mathrm{MeV}}

\newcommand{\fm}{\mathrm{fm}}

\newcommand{\chipt}{\chi\rm{PT}}

\newcommand{\beq}{\begin{equation}}   
\newcommand{\eeq}{\end{equation}}   
\newcommand{\beqn}{\begin{eqnarray}}  
\newcommand{\eeqn}{\end{eqnarray}}   
\newcommand{\nn}{\nonumber}

\def\mcC{{\mathcal C}}
\def\mcD{{\mathcal D}}
\def\mcJ{{\mathcal J}}

\def\mcO{{\mathcal O}}

\def\la{\langle}
\def\ra{\rangle}
\def\psibar{\overline{\psi}}
\def\chibar{\overline{\chi}}

\newcommand{\old}[1]{}

\newcommand{\plotangle}{0}

\begin{document}

\begin{flushright}
FTUAM-11-64\\
IFT-UAM/CSIC-11-94\\
DESY 12-021\\
SFB/CPP-12-07\\
ROM2F/2012/01\\
\end{flushright}

\begin{frontmatter}

\title{Sigma terms and strangeness content \\  of the nucleon with $N_f=2+1+1$  
\\twisted mass fermions}

\begin{center}
\includegraphics[width=100pt]{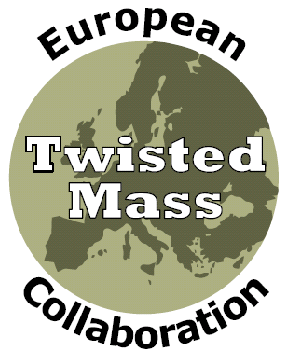}
\end{center}\vspace{10pt}
\author[DESY]{Simon Dinter},
\author[DESY]{Vincent Drach},
\author[RMII]{Roberto Frezzotti},
\author[MADRID]{Gregorio Herdoiza},
\author[DESY,RMII]{Karl Jansen},
\author[RMII]{Giancarlo Rossi}
\address[DESY]{NIC, DESY, Platanenallee 6, D-15738 Zeuthen, Germany}
\address[RMII]{Dip. di Fisica, Universit{\`a} di Roma Tor Vergata and \\ INFN Sez. di Roma Tor Vergata, Via della Ricerca Scientifica 1, I-00133 Roma, Italy}
\address[MADRID]{ Departamento de F\'isica Te\'orica and Instituto de F\'isica Te\'orica UAM/CSIC,\\
Universidad Aut\'onoma de Madrid, Cantoblanco, E-28049 Madrid, Spain }

\begin{abstract}

We study the nucleon matrix elements of the quark scalar-density operator 
using maximally twisted mass fermions with dynamical light ($u$,$d$), strange and charm 
degrees of freedom. We demonstrate that in this setup the nucleon matrix elements
of the light and strange quark densities can be obtained 
with good statistical accuracy, while for the charm quark counterpart only 
a bound can be provided. 
The present calculation which is performed at only one value of the lattice 
spacing and pion mass serves as a benchmark  
for a future more systematic computation of the scalar quark 
content of the nucleon. 
\end{abstract}

\begin{keyword}
strange content of the nucleon, nucleon sigma terms, lattice QCD
\end{keyword}

\end{frontmatter}

\section{Introduction}

A number of experiments (see for instance 
refs.~\cite{Ahmed:2010wy,Bernabei:2008yi,Aalseth:2010vx,Sumner:2010zz,Jochum:2011zz}) 
have been designed to investigate the nature of Dark Matter (DM) by 
detecting and/or measuring the recoil energy of nuclei hit by a hypothetical 
DM particle. One very popular example of such DM candidates is a  
weakly interacting massive particle (WIMP) like the ones that are predicted in 
a large class of models~\cite{Bertone:2004pz}.
However, definite evidences for a direct detection of WIMPs have not
   been observed up till now. Nonetheless, the various ongoing experiments provide  rather severe constraints for the parameters of many DM models. 

\begin{figure}[tb]
\begin{center}
\includegraphics[width=190pt,angle=\plotangle]{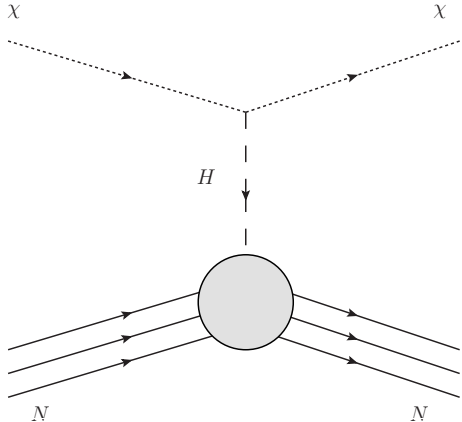}
\end{center}
\caption{The Higgs-boson exchange contribution to the WIMP-Nucleon low energy 
scattering process.}\label{fig:wimp_N_scatt}
\end{figure}

A possible scenario for the detection of a WIMP type of DM particles relies on the idea  
that the WIMP -- due to its assumed large mass -- produces a Higgs boson 
which in turn couples to the various quark flavour scalar density operators taken
between nucleon states, as depicted in Fig.~\ref{fig:wimp_N_scatt}. 
In fact, at zero momentum transfer, 
the cross section for spin independent (SI) elastic WIMP--nucleon ($\chi N$) 
scattering reads \cite{Ellis:2008hf}

\be
\sigma_{\rm SI, \chi N} \sim  
\Big\lvert\sum_{f} G_{q_f}(m_\chi^2 ) f_{T_f} \Big\rvert^2  \gap \text{with} \gap f_{T_f} 
= \frac{m_{q_f}}{m_N} \la N | \bar{q}_f{q_f} |N \ra \, .
\label{eq:crosssection} 
\ee

The functions $G_{q_f}$ depend on several parameters of 
the particular model $\sigma_{\rm SI, \chi N}$ is computed in, among which
the WIMP mass, $m_\chi$. A precise expression of the cross section in the constrained minimal supersymmetric Standard Model (CMSSM) can be inferred from 
ref.~\cite{Ellis:2008hf}. The expression above for $\sigma_{\rm SI, \chi N}$ is in particular valid in 
the SU(2) isospin limit for $u$ and $d$ quarks, an approximation that is always 
understood in the following.
What we want to compute is the magnitude of the dimensionless and renormalization group 
invariant (RGI) coupling $f_{T_f}$, which depends on the mass $m_{q_f}$ of the quark 
of flavour $f$ and the nucleon mass, $m_N$. For short in the following we will 
refer to the matrix  element $\la N | \bar{q}_fq_f |N \ra$ as the ``scalar $f$-quark content'' of the nucleon.

As one can see from Eq.~(\ref{eq:crosssection}), the cross section, $\sigma_{\rm SI, \chi N}$ depends quadratically on $f_{T_f}$, and it is thus very sensitive to the size of the scalar content contributions of different flavours. Already a O(10\%) variation
of $f_{T_f}$ can lead to significant changes in $\sigma_{\rm SI, \chi N}$. 
It is therefore necessary to compute accurately and with controlled error 
the hadronic matrix elements $\la N | \bar{q}_fq_f |N \ra$.   The implications of the hadronic uncertainty in comparing models with present data of WIMP direct detection has been for instance presented recently  in  \cite{Fornengo:2010mk}.     

One way to calculate for various flavours the scalar quark content of the nucleon  is provided by chiral perturbation theory ($\chipt$).   
The results of such a calculation are usually 
parametrized in terms of $\sigma_{\pi N}$ and $\sigma_0$ 
defined by the formulae  
\be\label{eq:sigma_terms}
\sigma_{\pi N} \equiv m_l \la N|  \bar{u} u+  \bar{d} d |N \ra \gap\textmd{and}\gap 
\sigma_0  \equiv  m_l \la N|  \bar{u} u+ \bar{d} d  - 2 \bar{s}s|N \ra\, ,
\ee
where $m_l$ is the common mass of the $u$ and $d$ light quarks. 
A low energy theorem relates $\sigma_{\pi N}$ to the pion nucleon 
scattering amplitude extrapolated to the Cheng-Dashen point~\cite{Cheng:1970mx}. 
The functional form of the extrapolation formula can be established using 
dispersion relations and in this way  
$\sigma_{\pi N}$ has been found to be 
$\sigma_{\pi N} = 79(7)~\mev$~\cite{Pavan:2001wz}. Note that a recent result which uses Lorentz covariant baryon chiral perturbation theory gives $\sigma_{\pi N} = 59(7)~\mev$\cite{Alarcon:2011zs}.
The non-singlet matrix element $\sigma_0$ can be obtained by looking at the pattern 
of the $SU(3)$ symmetry breaking visible in the spectrum of the baryon octet. 
A  determination using $\chipt$  obtained in ~\cite{Borasoy:1996bx} has given  $\sigma_0 =  36(7)\mev$.

A direct measure of the magnitude of the strange quark content of the nucleon 
relative to the light quark content is represented by the ratio
\be\label{eq:y_para}
y_N \equiv  \frac{ 2\la N| \bar{s} s |N \ra }{ \la N| \bar{u} u+ \bar{d} d |N \ra}\, .
\ee
The $y_N$-parameter can be related to $\sigma_{\pi N}$ and $\sigma_0$ by 
$y_N=1-\sigma_0/\sigma_{\pi N}$ and one finds,  
using the previously quoted numbers, $y_N = 0.44(13)$~\cite{Young:2009ps}.
This result leads to a surprisingly large strange quark content of 
the nucleon, and consequently to a large strange quark contribution to 
the $\sigma_{\rm SI, \chi N}$ cross section in \eq{eq:crosssection}.
On the other hand, the quoted uncertainty on $y_N$ is rather large, 
i.e.\ of the order of 30\%. As pointed out in ref.~\cite{Giedt:2009mr} 
and discussed above, it is quite important to provide a 
precise value for the strange quark content of the nucleon in order to be able 
to interpret the ongoing and planned experimental searches of WIMPs.

Lattice QCD can provide a determination of these nucleon matrix elements from first principles. The difficulty involved in these computations has limited for a long time the possibility to isolate a physical signal for these quantities. The aim of this paper is to show that it is indeed possible to accurately compute $y_N$ 
and test whether the strange quark content of the nucleon is as
large as indicated by $\chipt$. 
To this end, we will use Wilson lattice QCD with maximally twisted
mass fermions since this framework is particularly suited for such a calculation, 
as we will discuss below. 

On the lattice, two approaches have been followed for the calculation of the 
light and strange quark content 
of the nucleons. The first is based on the 
Feynman-Hellman theorem~\cite{Feynman:1939zz} that relates the nucleon scalar matrix 
element to the dependence of the nucleon mass on the quark masses via the equations 
\ba
&&\sigma_{\pi N} = m_l \la N|  \bar{u} u+  \bar{d} d |N \ra = m_l \frac{\partial m_N}{\partial m_l} \label{eq:sigma_terms_FH1}\, ,\\
&&\sigma_{sN} = m_s \la N| \bar{s} s |N \ra = m_s \frac{\partial m_N}{\partial m_s}\label{eq:sigma_terms_FH2}\, ,
\ea
where we denote by $m_s$ the strange quark mass and the derivative are intended to be evaluated at the quark mass values that correspond to the physical pion and kaon masses.
This approach is often referred to as the ``spectrum method''. 
The other approach, the so-called ``direct method'', consists in evaluating 
directly the matrix elements appearing in~\eqs{eq:sigma_terms_FH1} and~(\ref{eq:sigma_terms_FH2}).

Both approaches are numerically very challenging. In the spectrum method, a number 
of simulations at different values of the strange quark mass are needed in order to evaluate the derivative.  
In the direct method, the computation of the disconnected diagrams is required, 
which is a highly demanding task since they often show quite a bad signal to noise 
ratio, thus requiring very high statistics. In addition, 
as pointed out in ref.~\cite{Michael:2001bv}, when lattice 
discretizations that break chiral symmetry are used, a  
mixing between the bare light and strange scalar quark density matrix elements occurs 
under renormalization. This mixing is not present for chiral invariant (e.g.\ overlap) 
fermions~\cite{Ohki:2009mt}. It can also be avoided up to O($a^2$) effects when -- like we do in this work -- maximally twisted mass fermions are used 
in a mixed action setup as explained below.  

There exist already a number of lattice QCD computations of the strange quark 
content of the nucleon, see the works of 
refs.~\cite{Michael:2001bv,Ohki:2008ff,Bali:2009dz,Freeman:2009pu,Takeda:2010cw,Young:2009zb,Collins:2010gr,Durr:2010ni,Durr:2011mp,Bali:2011ks,Bali:2011rs,Horsley:2011wr} 
and the review paper~\cite{Young:2009ps}.
Although these calculations indicate that the strange quark content of the 
nucleon is smaller than suggested from 
$\chipt$, the quoted results are affected by large statistical errors, 
making it difficult to reach definite conclusions. 

In this paper we present a method which allows to compute the strange quark
content of the nucleon with small statistical errors. By means of a benchmark
calculation at only one value of the lattice spacing, volume and quark mass we
demonstrate that it is indeed possible to achieve for the parameters $f_{T_s}$
and $y_N$ (see Eqs.~(\ref{eq:crosssection}) and~(\ref{eq:y_para})) a signal
to noise ratio significantly (more than 4 standard deviations) 
different from zero.  It remains of course open the question of the
size of systematic effects which we plan to address in the future by including 
in the computation more lattice spacings, volumes and quark masses.

As we said, a main ingredient that enables us to obtain accurate values 
for the strange quark content of the nucleon is the use of Wilson lattice 
QCD with (maximally) twisted mass fermions for our simulations. 
Besides its property of automatic $O(a)$-improvement, it offers the advantage 
that special techniques for computing disconnected diagrams (see below)
can be employed, which significantly reduce the 
size of the numerical noise typically affecting the computation of such diagrams. 
In addition, with our choice of twisting for valence fermions
the operator matrix elements relevant for the various flavour contents of the
nucleon turn out to be all multiplicatively renormalizable. 
Although the same property is valid in chiral invariant (e.g.\ overlap)
lattice formulations, twisted
mass fermions are computationally much less demanding, enabling us to work 
with a  large number of gauge configurations and at fairly big volumes 
and small lattice spacing. 

As a last point, we want to mention that for this computation we employ gauge 
configurations generated with $N_f=2+1+1$ dynamical quarks, in which besides a 
mass degenerate $u$ and $d$ light doublet also a mass non-degenerate strange $s$ 
and charm $c$ pair is present in the sea.
This will allow us for the first time to also study the charm quark content of the nucleon.

\section{Computational methods}

\subsection{Lattice action}

The lattice action used in our simulations includes 
as dynamical degrees of freedom,
besides the gluon field,  
a mass-degenerate light up and down 
quark doublet as well as a strange-charm quark pair, a situation which 
we refer to as the $N_f=2+1+1$ setup. 
While in the pure gauge sector we use the Iwasaki action~\cite{Iwasaki:1985we}, 
for the fermion part twisted mass fermions are used.
In particular, concerning sea quarks, we make use of the formulation of
refs.~ \cite{Frezzotti:2000nk,Frezzotti:2003ni} for the light mass degenerate
$u$--$d$ sector, while the action introduced in
refs.~\cite{Frezzotti:2003xj,Frezzotti:2004wz} is employed
for the mass non-degenerate $c$--$s$ sector.
The quark mass parameters of the heavy flavour pair have
been tuned so that in the unitary lattice setup the Kaon and D-meson masses, 
take (approximatively) their 
experimental values. More information about this scheme and further simulation 
details can be found in ref.~\cite{Baron:2010bv}.

In order to fix the notation, we give here the explicit form of the twisted mass action for a doublet of degenerate quarks :
\ba
\hspace{-.8cm}&&S_f[\chi_f,\chibar_f,U] 
=  a^4 \sum_{x} \chibar_f(x)D_{\rm tm}^{(f)}[U]  \chi_f(x)
=  a^3 \sum_{x} \Bigg\{ \chibar_f(x)(\frac{1}{2\kappa_f}+  i  a\mu_f \gamma_5 \tau^3 )\chi_f(x) 
\nonumber \\
\hspace{-.8cm}&& - \chibar_f(x) \sum_{\mu=0}^3  
\Big[ U_{\mu}(x) \frac{1-\gamma_{\mu}}{2} \chi_f(x + a\hat{\mu}) 
+ U^{\dagger}_{\mu}(x-a\hat{\mu}) \frac{1 + \gamma_{\mu}}{2} 
\chi_f(x- a\hat{\mu})\Big] \Bigg\} \, . 
\label{eq:twba-2f-action}
\ea
Here $\mu_f > 0$ denotes the bare twisted mass. The hopping parameter $\kappa_f$ is an
alias for the bare standard mass $m_{0f}= ((2\kappa_f)^{-1} -4)/a$. Eq.~\ref{eq:twba-2f-action} defines $D_{\rm tm}^{(f)}[U]$ the flavour degenerate  Wilson twisted mass operator.
When $\kappa_f$  is tuned to its critical value, 
$\kappa_{\rm cr}$, the lattice QCD formulation known as maximally twisted mass 
fermions is achieved, which guarantees O$(a)$ improvement of physical observables. 
The value of $\kappa_{\rm cr}$ for all valence flavours is taken 
to be the same as in the sea quark sector~\cite{Frezzotti:2004wz}.

For further needs we also introduce the operators $D_{f,\pm}$ denoting the 
upper and lower flavour components of  $D_{\rm tm}^{(f)}[U] $, referred to as the Osterwalder-Seiler Dirac operator :
\be
D_{f,\pm}[U] = \rm{tr}~\left\{\frac{1\pm\tau_3}{2}D_{\rm tm}^{(f)}[U] \right\} ,
\label{eq:Dpm}
\ee
where $\rm{tr}$ denotes the trace in flavour space. $D_{f,\pm}[U]$ then is the Dirac operator of an Osterwalder-Seiler lattice quark with mass $\pm \mu_f$.

When we discuss below the 2-point and 3-point correlation functions necessary for this work, 
we will use the so-called physical basis of quark fields denoted as $\psi_f$. 
The physical field basis is related to the twisted quark field basis, $\chi_f$, by 
the following field rotation\footnote{At maximal twist the quark fields $\psi_f$, $\psibar_f$ are said to be in the ``physical'' quark basis if the quark mass term in the Lagrangian appears in the canonical form $\psibar_f \mu_f \psi_f$.}

\be\label{eq:rotation_phys}
\psi_f \equiv e^{i\frac{\omega_f}{2} \gamma_5 \tau^3} \chi_f \gap\textmd{and}\gap  \psibar_f \equiv \chibar_f  e^{i\frac{\omega_f}{2} \gamma_5 \tau^3},
\ee
where the twist angle $\omega_f=\pi/2$ at maximal twist. 
In the following, $\psi_f$ with index $f=l,s,c$ will denote  quark field doublets  of light ($l$), strange ($s$) or charm ($c$) quarks depending on the mass $\mu_f$ chosen in the valence sector. Since $\psi_f$ will always refer to the physical basis we will denote  with $u$ and $d$ the two components of $\psi_l$. Staying close to the notation of \eq{eq:Dpm} we will denote with $s_\pm$ (resp. $c_\pm$) the two components of $\psi_s$ (resp. $\psi_c$). 

In order to have a consistent mixed action setup, the values of the bare quark mass
parameters $\mu_s$ and $\mu_c$ in \eq{eq:twba-2f-action} have been tuned such 
that the Kaon and D-meson masses of the unitary setup\cite{Baron:2010bv} are matched.

In particular the matrix element entering $\sigma_{\pi N}$ will be calculated in such a way that (Euclidean)
unitarity is preserved at finite lattice spacing, while those of interest
for the strange and charm content of the nucleon are evaluated in a mixed action setup
where unitarity violations represent mere O($a^2$) artefacts~\cite{Frezzotti:2004wz}. 
Such cutoff effects are expected to be 
numerically small in line with the findings from previous mixed action studies carried out on ETMC 
$N_f=2$\cite{Constantinou:2010qv} and $N_f=2+1+1$ \cite{Farchioni:2010tb} gauge ensembles.

\subsection{Nucleon scalar matrix elements}
\label{eq:Nucleon_scalar}
The nucleon two-point function is defined in the physical quark basis by
\be \label{eq:C2pts}
C^{\pm }_{N,\rm 2pt}(\tau) =   \sum_{\vec{x}}  {\rm{tr}}~\left\{
\Gamma^{\pm}  \la \mcJ_{N}(x) \overline{\mcJ_{N}}(x_{\rm src})  \ra \right\} ,
\ee
where $\langle ... \rangle$ denotes 
field average,  $\Gamma^{\pm} = (1\pm\gamma_0)/{2}$ are the parity projectors,
$x_{\rm src}\equiv (t_{\rm src},\vec{x}_{\rm src})$ is the space-time location of the source
and $\tau\equiv t-t_{\rm src}$ stands for the source-sink  separation.
The subscript $N$ refers to the proton or to the neutron state for which the
interpolating fields are given by the formulae
\be
\mcJ_p = \epsilon^{abc} \left( u^{a,T} \mcC\gamma_5 d^b \right) u^c \gap\textmd{and}\gap   \mcJ_n = \epsilon^{abc} \left( d^{a,T} \mcC\gamma_5 u^b \right) d^c,
\label{eq:N-fields}
\ee
where $\mcC$ is the 
charge conjugation matrix. Note that due to translational invariance 
$C^{\pm }_{N,\rm 2pts}(\tau)$ does not depend on the spatial source 
location, $\vec{x}_{\rm src}$, which we can thus choose freely. 
Let us also 
recall that, since here we work in the $SU(2)$ isospin limit approximation, 
an exact symmetry of the action (i.e.\ ${\cal P}\times (u\leftrightarrow d$) 
where ${\cal P}$ is parity) leads to the relation     
$C^{\pm}_{n,\rm 2pt} (\tau) =  C^{\pm}_{p,\rm 2pt} (\tau)$~\cite{Alexandrou:2008tn}. 

The three-point functions of interest in this paper are defined by 
\be\label{eq:C3pts}
C^{\pm,f}_{N,\rm 3pt}(\tau,\tau_{\rm{op}}) =     
\sum_{\vec{x} ,\vec{x}_{\rm op}}   {\rm{tr}}~ ~\left\{
\Gamma^{\pm}\la \mcJ_{N}(x) O_f(x_{\rm op}) \overline{\mcJ_{N}}(x_{\rm src})  \ra\right\} \, , 
\ee

where  $O_f$ for $f=l,s,c$ denotes an appropriate 
(see below \eq{eq:ops_scalar}) 
lattice regularization of the light, strange or charm quark scalar density and
$\tau_{\rm op} =t _{\rm{op}} - t_{\rm{src}} $ is the operator-to-source time separation.
Since we are considering an operator with a non vanishing vacuum expectation value, 
we also need to introduce the corresponding vacuum subtracted correlator
\be\label{eq:C3pts_vev_sub}
C^{\pm,f,\rm{sub}}_{N,\rm 3pt}(\tau,\tau_{\rm{op}}) =           
C^{\pm,f}_{N,\rm 3pt}(\tau,\tau_{\rm{op}}) -      
C^{\pm }_{N,\rm 2pt}(\tau)   \sum_{\vec{x}_{\rm op}} \la O_f(x_{\rm op}) \ra\, .
\ee

To be specific, the operator $O_f$ will be given for our case by
\be\label{eq:ops_scalar}
O_l = \bar{u}u + \bar{d}d,~O_s= \hf \left(\bar{s}_+s_+ + \bar{s}_-s_-\right) \gap\textmd{and}\gap O_c= \hf \left(\bar{c}_+c_+ + \bar{c}_-c_-\right),
\ee
depending on the quark flavour of interest. Using these operators, we shall obtain the multiplicatively renormalizable,  O($a$) improved matrix elements relevant for this paper.

For each flavour $f$, the bare  scalar matrix 
element at zero momentum transfer $\la N(p) | O_f (0) | N(p) \ra$ can be written 
in terms of an effective coupling bare constant $g_{S,f}$ and the nucleon spinor $u_N$, in the form 
\be
\la N(p) | O_f (0) | N(p) \ra = g_{S,f} \bar{u}_N(p) u_N(p)  \, .
\ee
Using the two- and three-point correlators of~\eqs{eq:C2pts} 
and~(\ref{eq:C3pts}), we build for $f=l,s,c$ the ratio
\be\label{eq:ratio_def}
R_{f}(\tau,\tau_{\rm op}) \equiv \frac{C^{+,f,\rm{sub}}_{N,\rm 3pt}(\tau,\tau_{\rm op})}{C^{+}_{N,\rm 2pt}(\tau) }  = g_{S,f} + \mbox{O}( e^{-\Delta M  |\tau_{\rm op}|}) +  \mbox{O}( e^{-\Delta M |\tau -\tau_{\rm op}|} )\, ,\ee
where $\Delta M$ is the mass gap between the lowest nucleon state and the 
first excited state with the same quantum numbers. 
One can thus extract from the asymptotic 
time behaviour of the various $R_{f}(\tau,\tau_{\rm op})$
the bare effective scalar couplings $g_{S,f}$, which are in turn
simply related to the nucleon sigma terms of interest. For instance, at maximal
twist the lattice regulated versions of Eqs.~(\ref{eq:sigma_terms_FH1}) and~(\ref{eq:sigma_terms_FH2})
will read
\be
\sigma_{\pi N}^{\rm Lat} = \mu_l g_{S_l} \, , \qquad 
\qquad \sigma_{sN}^{\rm Lat} = \mu_s g_{S_s} \, .
\label{eq:sigma-term-Lat}
\ee
The systematic errors 
$\mbox{O}(e^{-\Delta M|\tau-\tau_{\rm op}|})$ and $\mbox{O}(e^{-\Delta M |\tau_{\rm op}|})$
originating from the finiteness of the time separations 
$\tau -\tau_{\rm op}$ and $\tau_{\rm op}$ will be neglected in this work. However they can 
have a non-negligible impact on the evaluation of nucleon matrix elements, as shown for 
instance in~\cite{Dinter:2011sg}. We are therefore planning to address this problem 
in a forthcoming publication.

\subsection{Lattice discretization and evaluation of correlators}
\label{subsec:lat_eval}

The main aim of the paper is to study whether the improved
methods to compute disconnected diagrams as applicable 
for twisted mass fermions will indeed lead to a
calculation of the quark contents of the nucleon with 
significantly reduced errors compared to earlier works.
The analysis performed in this work concentrates 
therefore on one ensemble of a $32^3\times 64$ lattice volume with a lattice 
spacing of $a=0.0779(4)~\fm$ ($\beta=1.95$) where the error quoted is only statistical\cite{Baron:2011sf},  and a pion mass of approximately 
$390~\mev$ ($a\mu_l=0.0055)$. 

In order to improve the overlap between the ground 
state and the interpolating fields we use Gaussian smearing of the quark fields 
appearing in the interpolating fields. We also use APE smearing of the gauge 
links involved in the Gaussian smearing, following the same 
strategy as in \cite{Dinter:2011jt,Drach:2010hy}.

\begin{figure}[htb]
\centering \subfigure[\label{fig:contract_conn}]%
{\includegraphics[width=0.45\linewidth]{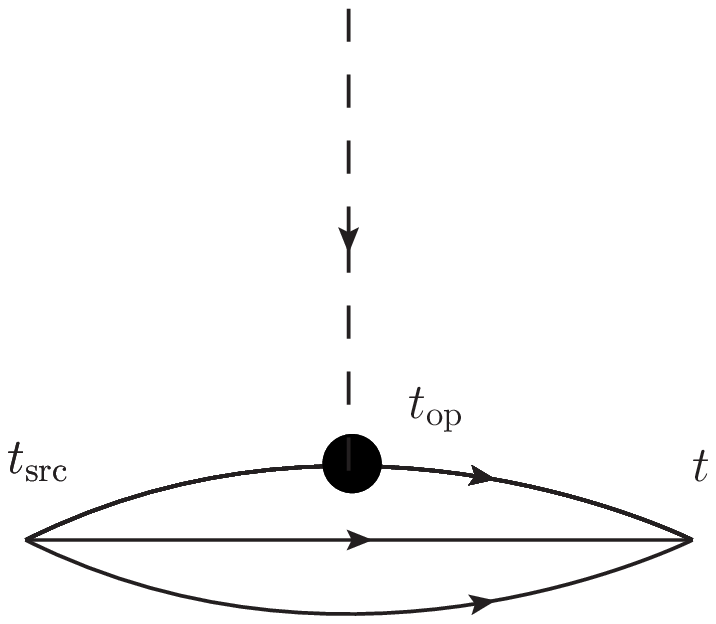}}
\quad \subfigure[\label{fig:contract_disc}]%
{\includegraphics[width=0.445\linewidth]{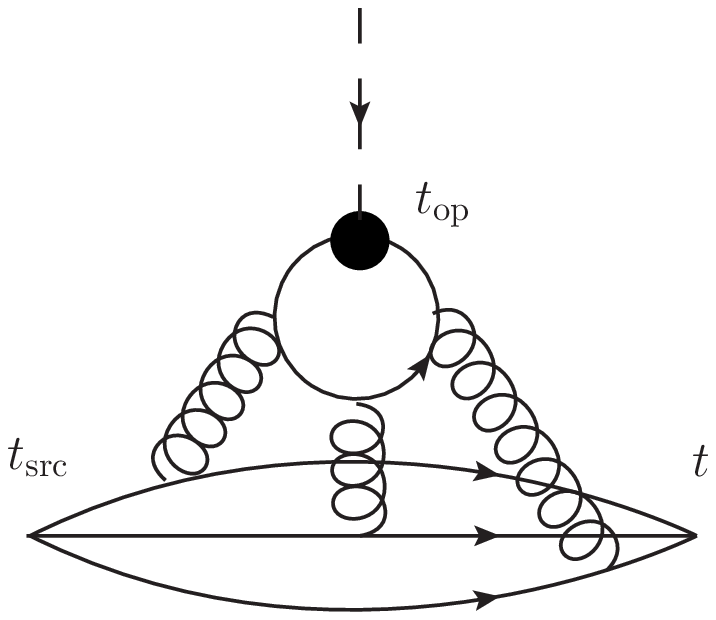}}
  \caption{Connected (left) and the disconnected
   (right) graphs arising from the Wick contractions of the 3-point function.}
  \label{fig:contract}
\end{figure}

In the twisted basis the scalar operators read
\be
\widetilde{O}_f=i\chibar_f\gamma_5 \tau^3 \chi_f,  \gap\textmd{where}\gap f=l,s,c~,
\ee
and are hence given by the pseudo scalar density.
There is a one-to-one correspondence 
between the bare operators $\widetilde{O}_f$ and the bare 
operators $O_f$ introduced in \eq{eq:ops_scalar} which is given 
at maximal twist by:
\be\label{eq:scalar_op_twisted_basis}
\widetilde{O}_f= i\chibar_f\gamma_5 \tau^3 \chi_f = 
\psibar_f \psi_f  =  \begin{cases}
~O_l & \textmd{if $f=l$}  \\ 
2 O_s & \textmd{if $f = s$}  \\ 
2 O_c  & \textmd{if $f = c$}  \\ 
  \end{cases}.
\ee
While the two-point nucleon correlators of \eq{eq:C2pts} give only rise to
quark-connected Wick contractions, in general the three-point functions of
\eq{eq:C3pts} yield both quark-connected (illustrated in Fig.~\ref{fig:contract}a)
and quark-disconnected (illustrated in Fig.~\ref{fig:contract}b) contributions.
In the following we will refer to them simply as to connected and disconnected
 fermionic Wick contractions (or diagrams) and shall write
\be
C^{\pm,f}_{N,\rm 3pt}(\tau,\tau_{\rm{op}}) = 
\widetilde{C}^{\pm,f}_{N,\rm 3pt}(\tau,\tau_{\rm{op}}) 
+ \mcD^{\pm,f}_{N,\rm 3pt}(\tau,\tau_{\rm{op}}) 
\ee
with $\widetilde{C}^{\pm,f}_{N,\rm 3pt}$ (resp.\ $\mcD^{\pm,f}_{N,\rm 3pt}$)
corresponding to the connected (resp.\ disconnected) quark diagrams, defined as
\ba
\hspace{-1.cm}&&\widetilde{C}^{\pm,f}_{N,\rm 3pt}(\tau,\tau_{\rm{op}}) = \!
\sum_{\vec{x} ,\, \vec{x}_{\rm op}} {\rm{tr}}~\left\{\Gamma^{\pm}\la  \big[ \mcJ_{N}(x) O_f(x_{\rm op}) \overline{\mcJ_{N}}(x_{\rm src})  \big]
\ra\right\} \, ,
\label{eq:C3pts-Wick} \\
\hspace{-1.cm}&&\mcD^{\pm,f}_{N,\rm 3pt}(\tau,\tau_{\rm{op}}) =\!
\sum_{\vec{x},\,\vec{x}_{\rm op}}  {\rm{tr}}~\left\{
\Gamma^{\pm}  \la 
\big[ \mcJ_{N}(x) \overline{\mcJ_{N}}(x_{\rm src}) \big]
\big[  O_f(x_{\rm op}) \big]
\ra\right\} \, ,
\label{eq:D3pts-Wick}
\ea
where the symbol $[...]$ is a shorthand for all the {\em connected} 
fermionic Wick contractions.
In particular, the contribution of the disconnected fermion loop 
to $\mcD^{\pm,f}_{N,\rm 3pt}$ on
a given gauge configuration $U$ 
in our setup reads
\be
\left[  O_f(x_{\rm op}) \right] = -i w_f \sum_{\vec{x}_{\rm op}} 
{\rm{tr}}~ ~\left\{\gamma_5 \left ( \frac{1}{D_{\rm tm}^{(f,+)}[U]}  -
 \frac{1}{D_{\rm tm}^{(f,-)}[U]}  \right )_{(x_{\rm op},x_{\rm op})}~\right\}   \, , 
\,\,f=l,s,c \, , \label{eq:disc_part}
\ee
where, in view of \eq{eq:scalar_op_twisted_basis}, 
$w_l=1$, $w_s=w_c=1/2$ and $D_{\rm tm}^{(f,\pm)}[U]$ are the Osterwalder-Seiler Dirac operators defined in  Eq.~(\ref{eq:Dpm}). 

For the strange and charm content of the nucleon only disconnected diagrams contribute
to the three-point correlator, while for the light quark content both kinds of fermionic
diagrams matter. The connected contributions to 
$\langle N(p)|O_l|N(p)\rangle$ have been evaluated using standard techniques for
three-point functions (``sequential inversions through the sink''). 
In this method one needs to fix the sink-to-source separation 
$\tau = t-t_{\rm src}$ and we choose, as in 
ref.~\cite{Dinter:2011jt}, $\tau = 12a$ 
corresponding in physical units to a separation of $\tau \approx 1~\fm$.

Since, using discrete symmetries and anti-periodic boundary 
conditions in the time direction for the quark fields, one finds
\bea
&&C^{+}_{N,\rm 2pt}(\tau)  = -C^{-}_{N,\rm 2pt}(T-\tau) \label{eq:time-symm-prop1}\\
&&C^{+,f}_{N,\rm 3pt}(\tau,\tau_{\rm{op}}) = - C^{-,f}_{N,\rm 3pt}(T-\tau,T-\tau_{\rm{op}}) \, , \label{eq:time-symm-prop2}
\eea
 where  $T$ denotes the lattice time extent, in order to increase the signal over noise ratio we have averaged
contributions related by the symmetry relations~(\ref{eq:time-symm-prop1}) and~(\ref{eq:time-symm-prop2}).
In addition, we have carried out Dirac matrix inversions at a number 
(denoted by $N_{\rm src}$ in the following) of
randomly chosen source points per gauge configuration, with the goal of better
exploiting the gauge field information contained in each configuration.  

An important issue to be discussed is renormalization  of the correlators introduced in sect. \ref{subsec:lat_eval}.  The technical arguments are given in Appendix~A. We summarize here the conclusions.
After having subtracted the mixing with the identity in the correlation function (see \eq{eq:C3pts_vev_sub}), the operator $O_l,O_s$ and $O_c$ do not mix among each other. Since also the bare quark mass $\mu_f$ renormalizes multiplicatively with a renormalization constant that is precisely the inverse of the one occurring in the renormalization of $O_{f}$, the lattice quantities $\mu_f g_{S_f} $, for $f=l,s,c$,  yield  O($a$) improved renormalization group invariant (RGI) sigma terms (see \eq{eq:sigma-term-Lat}). The renormalization pattern is thus as straightforward  as for chirally invariant overlap fermions.

\subsection{Numerical estimate of disconnected loops}
\label{subsec:vv_method}

Let us shortly sketch the variance reduction method for the evaluation of
 $$ {\rm{tr}}~\left\{ \Gamma 
 \left ( \frac{1}{D_{\rm tm}^{(f,+)}[U]}  -
 \frac{1}{D_{\rm tm}^{(f,-)}[U]} \right)_{(x,x)}\right\} \, ,
 \qquad \Gamma = \; {\rm some} \; {\rm Dirac} \; {\rm matrix} \, , $$
with twisted mass 
fermions introduced in~\cite{Michael:2007vn,Boucaud:2008xu}. The method
has already been applied to study the $\eta'$ meson in~\cite{Jansen:2008wv}.
It relies on the fact that the difference between the twisted basis Dirac 
matrices $D_{\rm tm}^{(f,+)}$ and $D_{\rm tm}^{(f,-)}$ is proportional to 
the identity ,
\be
D_{\rm tm}^{(f,+)} - D_{\rm tm}^{(f,-)} = 2 i  \mu_f  \gamma_5
\label{eq:diff_Dp_Dm}
\ee
implying that
\be\label{eq:vv_method}
\frac{1}{ D_{\rm tm}^{(f,-)} } - \frac{1}{ D_{\rm tm}^{(f,+)}  } = \frac{1}{ D_{\rm tm}^{(f,+)} }  
\left( D_{\rm tm}^{(f,+)} - D_{\rm tm}^{(f,-)} \right)  \frac{1}{ D_{\rm tm}^{(f,-)} } 
= 2 i   \mu_f \frac{1}{ D_{\rm tm}^{(f,+)} }  \gamma_5   \frac{1}{ D_{\rm tm}^{(f,-)} } \, .
\ee
For the practical calculation, we introduce a set $\Xi$ of $N_\xi$ independent 
random volume sources,   $\{\xi_{[1]},\dots, \xi_{[r]}, \dots, \xi_{[N_\xi]}\}$, satisfying 
\be
\lim_{N_\xi \to \infty}\left[ \xi^{  i}_{ [r]}(x)^\ast \xi^j_{ [r]}(y) \right]_\Xi 
= \delta_{xy} \delta^{ij} 
\label{eq:Rsources}
\ee
where $i=1,...,12$ refers to the spin and color indices of the 
source and $[ \dots]_\Xi$ denotes the average over the $N_\xi$
noise sources in $\Xi$ .  

Multiplying \eq{eq:vv_method} by a $\Gamma$ matrix and the noisy sources
$\xi_{[r]}^*(y)$ and $\xi_{[r]}(x)$ and  
taking the trace over spin and color indices we get 
\be\label{eq:vv_method_disc}
 2i   \mu_f \sum_{y} \left[ \phi_{[r]}^\dagger(y) \gamma_5 \Gamma  \phi_{[r]}(x) \right]_\Xi  
=  {\rm{tr}}~ \left\{ \Gamma \left( \frac{1}{ D_{\rm tm}^{(f,-)} } 
                           - \frac{1}{ D_{\rm tm}^{(f,+)} }\right)_{(x,x)}\right\}  + \mcO\left( N_\xi^{-1/2}\right),
\ee
where  
\be
 \phi_{[r]}=(1/D_{\rm tm}^{(f,+)}) \xi_{[r]} \gap\textmd{and}\gap 
\phi_{[r]}^\dagger = \xi_{[r]}^\dagger (1/D_{\rm tm}^{(f,+)})^{\dag} 
 = \xi_{[r]}^\dagger \gamma_5 (1/D_{\rm tm}^{(f,-)}) \gamma_5 \, .
\ee
For the generation of the random sources we have used
a $\mathbf{Z}_2$ noise taking
all field components randomly from the set $\{1,-1\}$.
We note that in the case of $\Gamma=\gamma_5$, the quantity in \eq{eq:vv_method_disc},
after summation over $\vec{x} \equiv \vec{x}_{\rm op}$,
provides an unbiased estimator of the 
disconnected fermion loops of \eq{eq:disc_part}.

\subsection{Performance of the method for disconnected loops}

As a first step, we performed a study 
to determine the optimal number, $N_\xi$, of
stochastic volume sources
to be used for evaluating the disconnected diagrams of \eq{eq:disc_part}.
For a given number $N_{\rm conf}$ of gauge field configurations increasing 
$N_\xi$ beyond some value will not improve the signal over noise ratio (SNR)
since the noise induced by the fluctuations of the gauge 
fields will eventually become dominating.

\begin{figure}[htb]
\begin{center}
\includegraphics[width=300pt,angle=\plotangle]{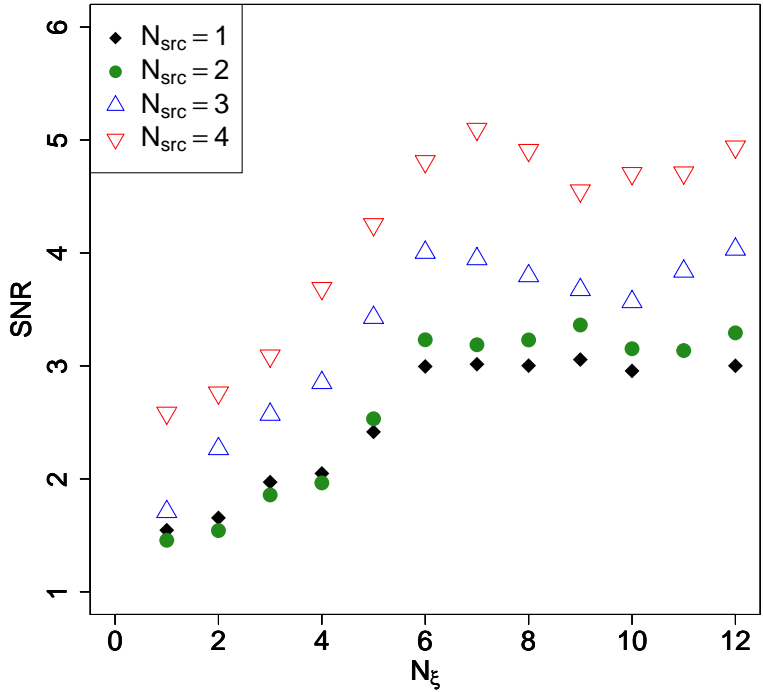}
\end{center}
\caption{Signal to noise ratio (SNR) of the quantity $R_{s}(\tau,\tau_{\rm op})$,
see Eq.~(\ref{eq:ratio_def}), for fixed 
values of $\tau/a=12$ and $\tau_{\rm op}/a=6$ as a function of $N_\xi$ and for different values of $N_{\rm src}$. 
The bare strange valence quark mass is  $a\mu_s = 0.016$. The number of
configurations used is $677$.  }\label{fig:scaling_R}
\end{figure}

For the present test we have used  $677$ gauge field configurations
and chose fixed time separations, 
$\tau=12a$ and $\tau_{\rm op} = 6a$. 
Fig.~\ref{fig:scaling_R} shows the SNR  
as a function of the number of stochastic source samples $N_\xi$
employed to evaluate the disconnected loops of \eq{eq:disc_part}.
As indicated in the figure, for the computation of the two point function we have used $N_{\rm src}=1,2,3,4$ randomly chosen point sources per gauge configuration.

The figure demonstrates    
that the signal over noise ratio reaches a plateau for 
$N_\xi>7$, meaning that for larger values of $N_\xi$ 
the error is dominated by the gauge field fluctuations. 
The finding that for $N_\xi>7$ a plateau of the SNR is reached
holds true for all values of $N_{\rm src}$ we have used. 
In addition we have checked that the above conclusion remains
essentially valid in the quark mass range we intend to explore.

Fig.~\ref{fig:scaling_R} also demonstrates that the SNR increases 
when more source points per gauge field configuration are used. 
When 
we change the number of source points 
from $1$ source per configuration to $4$, we find 
a decrease of the error by a factor of  
approximately $\sim 1.6$. Although this does not correspond
to the optimal factor of $2$, using $N_{\rm src}$-values moderately
larger than $1$ turns out to be
a convenient and economic way to increase the signal over noise ratio. In the final analysis, we will  always use $N_\xi=12$ and $N_{\rm src}=4$.

\begin{figure}[htb]
\begin{center}
\includegraphics[width=380pt,angle=\plotangle]{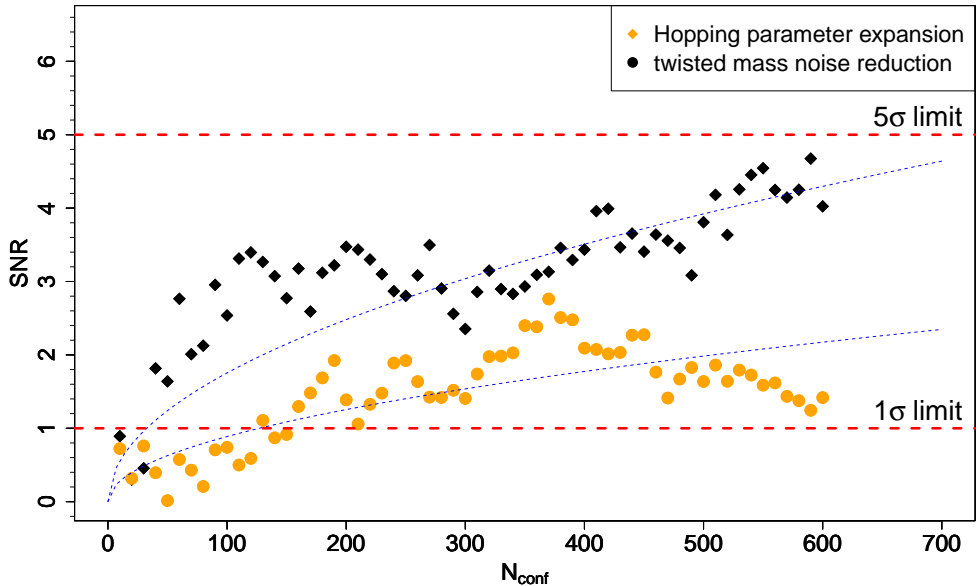}
\end{center}
\caption{Signal to noise ratio (SNR) of the quantity $R_{s}(12a,6a)$,
see Eq.~(\ref{eq:ratio_def}), 
for $N_\xi=12$ and $N_{\rm src}=4$ as a function of the number of gauge field configurations 
$N_{\rm{conf}}$, for the variance noise reduction 
technique used in this paper and the hopping parameter 
expansion technique. The dashed curves are drawn to guide the eye. 
The bare strange valence quark mass is the same as in \fig{fig:scaling_R}.  }
\label{fig:rel_err_scaling}
\end{figure}

In \fig{fig:rel_err_scaling} we compare the efficiency of the method 
discussed in sect.~\ref{subsec:vv_method}, which is based on the peculiar
property, see \eq{eq:diff_Dp_Dm}, of twisted mass fermions, with another noise reduction 
technique relying on the hopping parameter expansion of the Dirac operator. 
This latter technique is not restricted to twisted mass lattice QCD and  has been 
introduced in~\cite{Thron:1997iy,Michael:1999rs}. We refer the interested reader to  
the appendix B of~\cite{Boucaud:2008xu} for an implementation in the case
of twisted mass fermions. As can be seen from the figure the twisted mass specific 
variance reduction technique improves the signal over noise ratio by a factor $\sim 3$.
Performing a simple extrapolation in the 
number of gauge field configurations we estimate that 
with the hopping parameter expansion technique $\mcO(10000)$ configurations 
would be needed to reach a result $5\sigma$ away from zero, 
while only $\mcO(1000)$ configurations are necessary to obtain the same 
accuracy with our twisted mass specific technique.

\section{Results}

\subsection{The Pion-Nucleon $\sigma$-term, $\sigma_{\pi N}$}

\begin{figure}[htb]
\begin{center}
\includegraphics[width=300pt,angle=\plotangle]{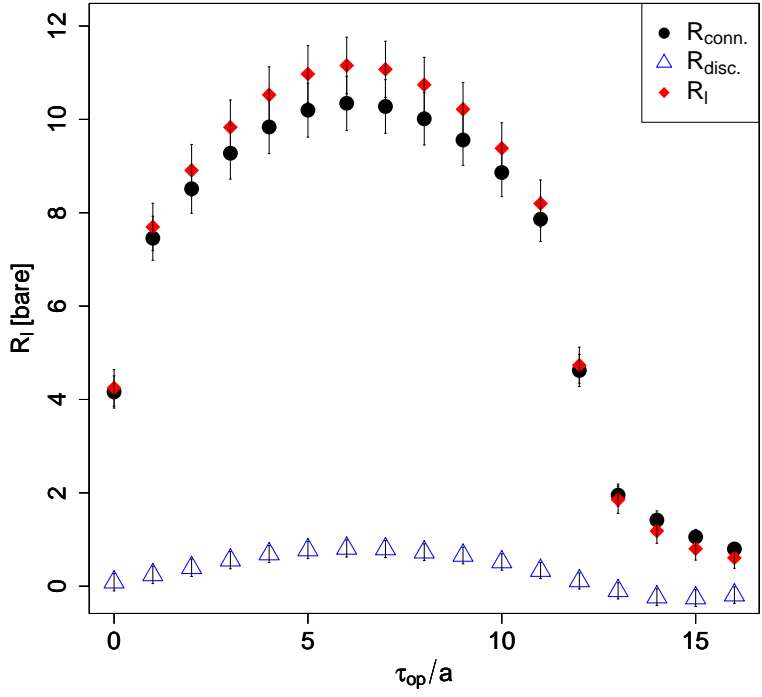}
\end{center}
\caption{Plot of the contributions $R_{\rm{disc.}}$ (blue triangle), $R_{\rm{conn.}}$ 
(black circles) and of their sum, $R_{l}$ (red diamonds) as function of $\tau_{op}$ at $\tau=12a$ for  $a\mu_l=0.0055$ and $\beta=1.95$.}
\label{fig:bare_light_plateau}
\end{figure}

We first concentrate on the determination of $\sigma_{\pi N}$ defined in \eq{eq:sigma_terms}. Since for this quantity only the up and down quarks come into play, 
we work in a fully unitary setup, where valence and sea quarks are regularized in the same way.
In the following we will denote by $R_{\rm{conn.}}$ (resp.\  $R_{\rm{disc.}}$) the contribution of $\widetilde{C}^{+,f}_{N,\rm 3pt}$  (resp.\ $\mcD^{+,f}_{N,\rm 3pt}$), see Eqs.~(\ref{eq:C3pts-Wick}, \ref{eq:D3pts-Wick}), to the ratio $R_l$ defined in \eq{eq:ratio_def}. In \fig{fig:bare_light_plateau} we show our results obtained for $R_{l}$, $R_{\rm{conn.}}$ and $R_{\rm{disc.}}$ as functions of $\tau_{op}=t_{op}-t_{src}$ for a fixed sink to source separation, $\tau=t-t_{src}=12a$. $R_{\rm{disc.}}$ has been computed using measurements over $842$ configurations with $N_\xi=12$ randomly chosen volume sources. $R_{\rm{conn.}}$ has been 
computed using $510$ configurations by employing the fixed sink method. 

The connected part, $R_{\rm{conn.}}$, 
denoted by the black filled circles, shows a pronounced time dependence indicating the  
contribution of excited states. In this work, we do not attempt to quantify 
the size of this systematic effect since our goal here is more to investigate 
whether statistically significant values for the scalar quark contents of the 
nucleon can be obtained.  
The disconnected part, $R_{\rm{disc.}}$, denoted by blue triangles 
in \fig{fig:bare_light_plateau}, clearly corresponds to a small contribution compared to the connected part $R_{\rm{conn.}}$, of the order of $\sim 10\%$ of the full ratio, $R_{l}$, represented by the red diamonds in the figure.  

\begin{figure}[htb]
\begin{center}
\includegraphics[width=300pt,angle=\plotangle]{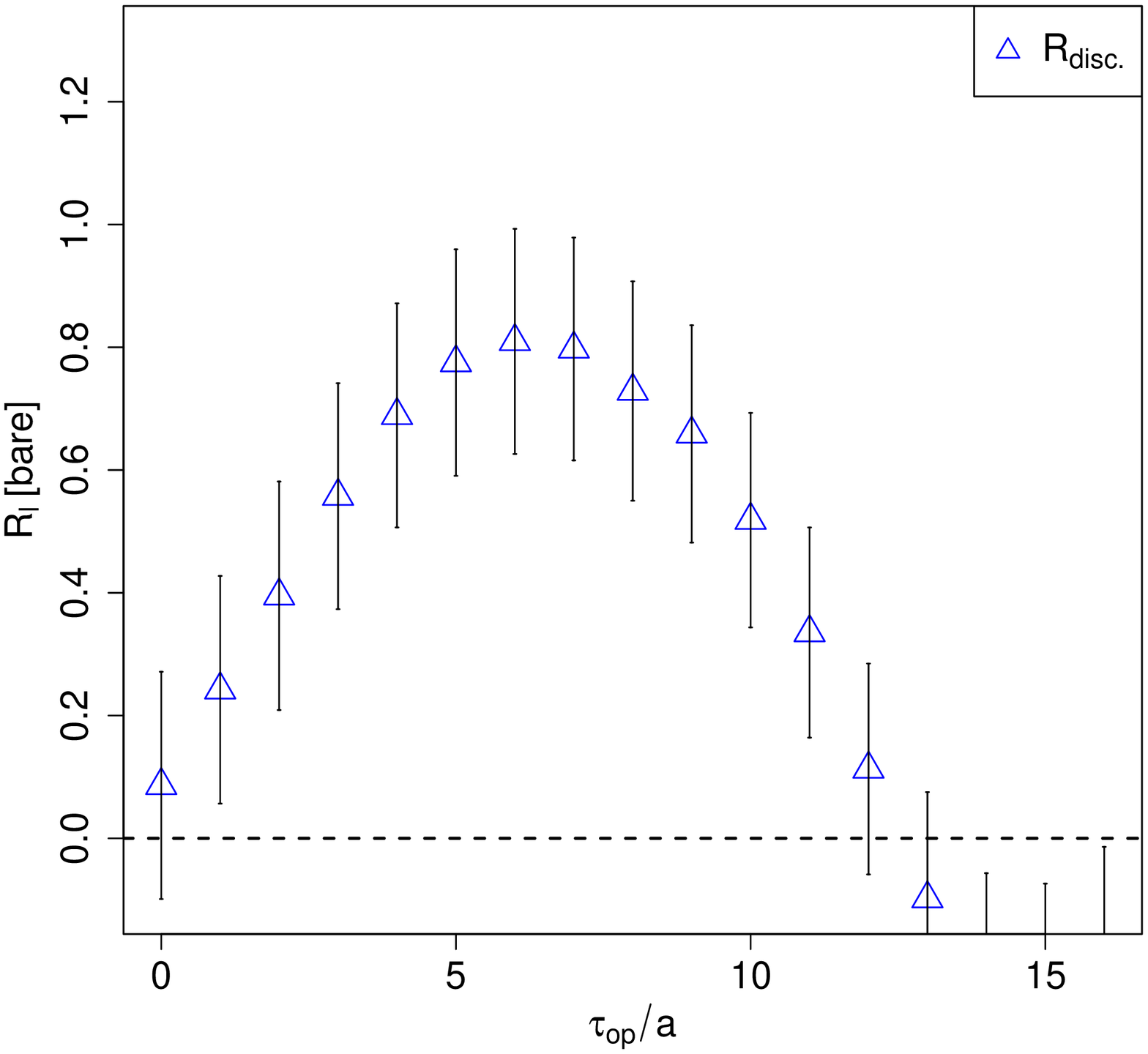}
\end{center}
\caption{Zoom of \fig{fig:bare_light_plateau} only showing $R_{\rm{disc.}}$ versus 
$\tau_{op}$ at $\tau=12a$.}\label{fig:disc_bare_light_plateau}
\end{figure}

In \fig{fig:disc_bare_light_plateau} (a zoom of \fig{fig:bare_light_plateau})     
we show only the disconnected contribution. Note  
the change in the scale on the vertical axis. 
It is encouraging that, by employing the techniques described above, 
we can indeed obtain a non-zero signal at a $\sim 4 \sigma$ level.                    
In order to determine the ``plateau'' values of $R_{\rm{conn.}}, R_{\rm{disc.}}$  
and $R_{l}$, we performed several fits to a constant through the data 
varying the fit interval. 
The results are summarized in \tab{tab:fit_sigma_piN}. 
We find that the disconnected contribution is about $\sim 8\% $ of the 
connected one. Nevertheless since the error on the connected 
contributions is smaller than the value of $R_{\rm{disc.}}$, the disconnected contribution 
cannot be neglected when computing the ratio $R_{l}$. 
We finally remark that all the statistical errors in this work 
are computed using the bootstrap method~\cite{EfronTibshirani:1993}.

\begin{table}[h]
\begin{center}
\begin{tabular}{|c|ccc|ccc|ccc|}
   \hline
    & \multicolumn{3}{c|}{ $R_{\rm{disc.}}$ ($842$ meas.)  } & \multicolumn{3}{c|}{ $R_{\rm{conn.}}$ ($510$ meas.)} &    \multicolumn{3}{c|}{$R_{l}$}  \\
   \hline
   $[\tau_{{op}_1},\tau_{{op}_2}]$ & fit & $\chi^2/\rm{ndof}$ & CL  & fit & $\chi^2/\rm{ndof}$ & CL & fit & $\chi^2/\rm{ndof}$ & CL \\
    \hline    
    $[2,10]$ & $0.66(16)$ &  $4/8$    & $0.85$ & $9.6(5)$  &  $12/8$   & $0.14$& $10.3(5)$ &  $15/8$   & $0.05$  \\
    $[3,9]$  & $0.72(17)$ &  $1/6$    & $0.98$ & $9.9(6)$  &  $3.2/6$  & $0.78$& $10.6(6)$ &  $4/6$    & $0.65$  \\
    $[4,8]$  & $0.76(18)$ &  $0.26/4$ & $0.99$ & $10.2(6)$ &  $0.6/4$  & $0.97$& $10.9(6)$ &  $0.75/4$ & $0.94$  \\
    $[5,7]$  & $0.79(18)$ &  $0.02/2$ & $0.99$ & $10.3(6)$ &  $0.04/2$ & $0.98$& $11.1(6)$ &  $0.05/2$ & $0.98$  \\ \hline
\end{tabular}
\caption{Plateau values for the ratio $R_{\rm{disc.}},R_{\rm{conn.}}$ and $R_{l}$ relevant  for the extraction of $\sigma_{\pi N}$ for different time intervals $[\tau_{{op}_1},\tau_{{op}_2}]$. We also include the $\chi^2$ by degrees of freedom ($\chi^2/\rm{ndof}$) and the confidence level (CL).}
\label{tab:fit_sigma_piN}
\end{center}
\end{table}

An estimate of the systematic error on $\sigma_{\pi N}$ can be given on the 
basis of the spread of the results one gets by varying the time interval  $[\tau_{{op}_1},\tau_
{{op}_2}]$ over which the plateau is taken, as displayed in \tab{tab:fit_sigma_piN}. For this purpose we 
construct the distribution of all fit results, weighted by their 
confidence level, and take the variance of this distribution as 
our estimate of the systematic errors. 
Using this procedure we find
\be
\sigma_{\pi N}(m_{\rm{PS}} \approx 390 \,\mev ) = 151 (8)(4)\, \mev \, , 
\ee
where the errors correspond to statistical and systematic 
uncertainties, respectively. Note that the dominant contribution to the systematic
error comes from the connected part of the ratio $R_{l}$.
Also, the value obtained here corresponds to only one pion mass of about $390~\mev$  and an extrapolation 
to the physical value of the pion mass will be finally needed.  
Notice that chiral perturbation theory 
predicts that $\sigma_{\pi N}$ vanishes in the zero quark mass limit.
While the calculation of $\sigma_{\pi N}$ at several quark masses and lattice spacings is beyond the scope of this paper, we remark that simulations in this directions are under way.

\subsection{Strange content of the nucleon}
\label{subsec:strange_content}

\begin{figure}[htb]
\begin{center}
\includegraphics[width=300pt,angle=\plotangle]{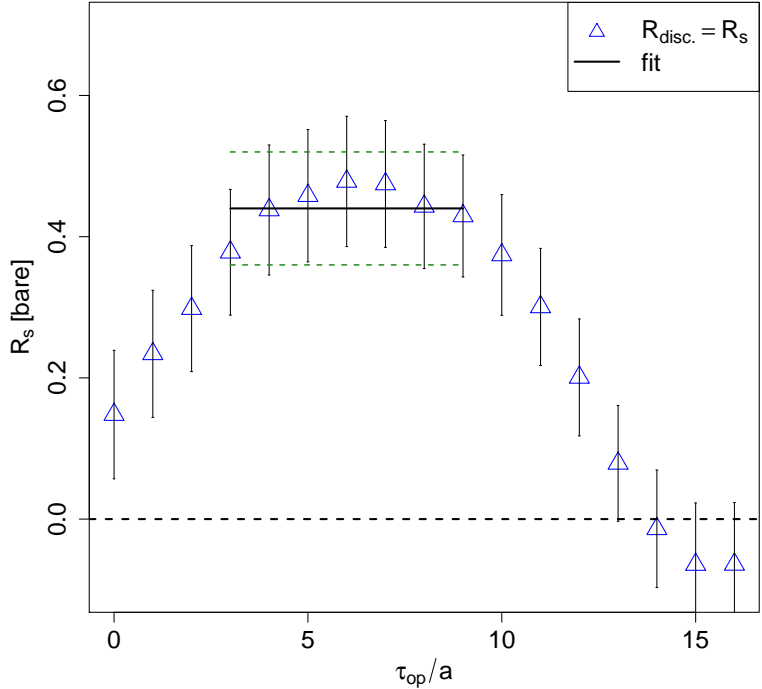}
\end{center}
\caption{Plot of $R_{\rm{disc.}}$, i.e. the bare ratio of \eq{eq:ratio_def}, 
versus $\tau_{op}$  at $\tau=12a$ in the strange quark mass regime ($a\mu_s=0.016$).}
\label{fig:plateau_strange}
\end{figure}

For each value of the valence  quark mass, one can define the quantity:
\be\label{eq:f_ts}
f_{T_f} = \frac{\mu_{f}}{m_N} \la N |O_{f} |N\ra = \frac{\mu_{f}g_{S,f}}{m_N} , \gap f=l,s,c \, ,
\ee
where $\mu_{f}$ is the bare quark mass and $\la N |O_{f} |N\ra$ the bare matrix 
element corresponding to the quark flavour $f$.  
As argued in Appendix~A, $f_{T_f}$ is a RGI quantity in our mixed action setup. 
For the ensemble used in this study, the nucleon mass $am_N$ has been
  determined in  \cite{Drach:2010hy} and is $am_N= 0.510(7)$. In \fig{fig:plateau_strange} we show  $R_{\rm{disc.}}$ versus $\tau_{op}$ 
for a quark mass of $a\mu_s = 0.016$. We see that the ratio is $\sim 5\sigma$ 
away from zero in the middle of the ``plateau''. 

We have performed several fits varying the time interval to extract a ``plateau'' value. 
Results for $f_{T_s}, \sigma_0$ and $y_N$ are summarized 
in \tab{tab:fit_strange_content}.  
While $y_N$ does not depend strongly on the fit window and seems thus to be, within our 
accuracy, free of excited state contaminations, $f_{T_s}$ and  $\sigma_0$ are affected by an excited state contamination in a similar way as we observed for $\sigma_{\pi N}$.
As we already remarked before, a quantitative evaluation of this systematic effect 
goes beyond the goal of this paper and will be addressed in the future. 
Our present results at a pion mass about $ 390 \,\mev$  for  $f_{T_s}, \sigma_0$ and $y_N$ are
\bea
&& f_{T_s}= 0.014(5)(1) \, ,\label{FT}\\
&& \sigma_{0}=137(7)(4)~\mev  \, ,\label{S0}\\
&& y_N=  0.082(16)(2) \label{YN}\, ,
\eea
where the first number in parenthesis represents the statistical error and the second 
the systematic uncertainty. In order to estimate the magnitude of systematic effects   
the same strategy as in the case of $\sigma_{\pi N}$ has been employed. The 
statistical error on $f_{T_s}$ also includes the error on the nucleon mass 
determination. Note that the value of $y_N$ quoted is obtained directly from the ratio of three point functions $C^{\pm,s,\rm{sub}}_{N,\rm 3pt}(\tau,\tau_{\rm{op}})$ and $C^{\pm,l,\rm{sub}}_{N,\rm 3pt}(\tau,\tau_{\rm{op}})$ and agrees within error with the value of $y_N$ estimated from $\sigma_0$ and $\sigma_{\pi N}$ ($y_N = 1 - \sigma_0/\sigma_{\pi N} \approx 0.092 $).  We stress again that our measurement of the strange content 
of the nucleon leads to a value of $y_N$ which is different from zero at a 
$5\sigma$ level. It is interesting to compare $f_{T_s}$ to the value, $f_{T_l}$, 
one gets in the light quark sector, for which we obtain a significantly larger value, 
namely $f_{T_l} =0.117(12)(3)$ (at $m_{PS}\sim 390~\mev$).      

\begin{table}[htb]
\begin{center}
\begin{tabular}{|c|ccc|ccc|ccc|}
   \hline
    & \multicolumn{3}{c|}{ $f_{T_s}$  } & \multicolumn{3}{c|}{ $\sigma_0$ ($\mev$)} &    \multicolumn{3}{c|}{ $y_N$-parameter}  \\
 \hline
   $[\tau_{{op}_1},\tau_{{op}_2}]$ & fit & $\chi^2/\rm{ndof}$ & CL  & fit & $\chi^2/\rm{ndof}$ & CL & fit & $\chi^2/\rm{ndof}$ & CL \\
 \hline
 $[2,10]$ & $0.013(2)$ &  $3/8$    & $0.90$ & $130(8)$  &  $11/8$   & $0.18$& $0.08(2)$ &  $0.7/8$ &  $0.99$  \\
 $[3,9]$  & $0.014(2)$ &  $0.8/6$  & $0.99$ & $135(8)$  &  $3/6$    & $0.79$& $0.08(2)$ &  $0.2/6$ &  $0.99$  \\
 $[4,8]$  & $0.014(3)$ &  $0.2/4$  & $0.99$ & $138(8)$  &  $0.5/4$  & $0.97$& $0.08(2)$ &  $0.03/4$ &  $0.99$  \\
 $[5,7]$  & $0.015(3)$ &  $0.03/2$ & $0.99$ & $140(8)$  &  $0.01/2$ & $0.98$& $0.09(2)$ &  $0.01/2$ &  $0.99$  \\
    \hline
\end{tabular}
\end{center}
\caption{Results of the fits to $f_{T_s}, \sigma_0$ and $y_N$ at $a\mu_s = 0.016$. The notations are the same as in Table~\ref{tab:fit_sigma_piN}. }
\label{tab:fit_strange_content}
\end{table}

\subsection{Charm quark content}

Following the same strategy as in sect. \ref{subsec:strange_content}, we have carried 
out the first study of the charm quark content of the nucleon. 
This is possible because we have at our disposal $N_f=2+1+1$ simulations with a fully 
dynamical charm quark degree of freedom. 

\begin{figure}[htb]
\begin{center}
\includegraphics[width=300pt,angle=\plotangle]{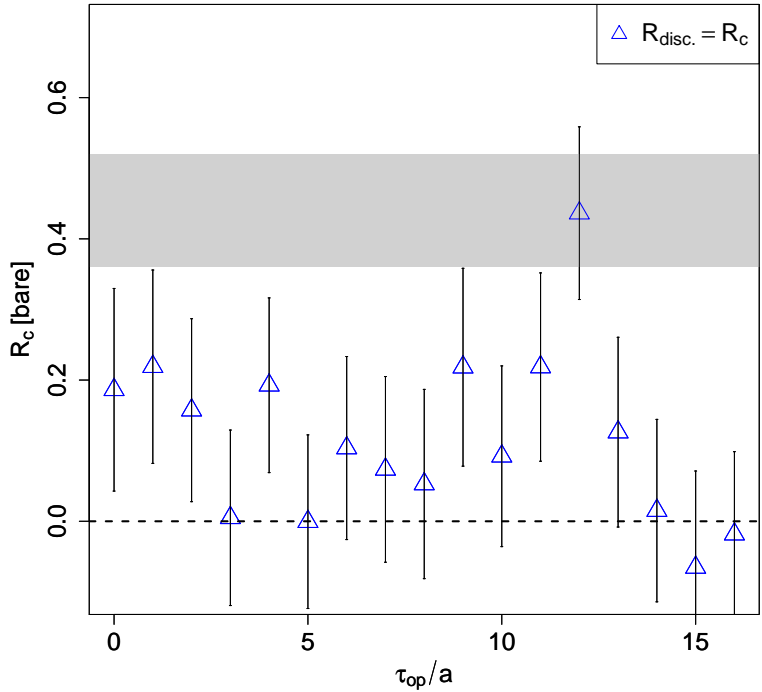}
\end{center}
\caption{The ratio $R_c$ of \eq{eq:ratio_def} for fixed $\tau=12a$ versus $\tau_{\rm{op}}$  in the charm quark mass 
regime ($a\mu_c=0.2$). For comparison, we display as a grey band the plateau value obtained 
in the strange quark case, see \fig{fig:plateau_strange}.}
\label{fig:plateau_charm}
\end{figure}

We show in \fig{fig:plateau_charm} the dependence of $R_{c}$ on $\tau_{op}$ 
(at $\tau=12a$), using exactly the same statistics 
as in the light and in strange quark sectors. 
Unfortunately, for the charm quark content no hint of a plateau is visible. Signal and noise have equal order of magnitude and our results are compatible with zero. 
For comparison we also show the results for the strange 
quark content obtained in the previous section as a grey band. From our data we can only establish the 
inequality 
\be
|\la N | O_c | N \ra|  \lesssim |\la N | O_s | N \ra| \, .
\ee

\section{Conclusion}

In this work we have performed a benchmark calculation of the 
scalar quark contents of the nucleon by directly computing the 
matrix elements $\la N | O_f |N \ra$ for $f=l,s,c$. 
Extending the calculation to strange and the charm quark flavours became 
possible owing to $N_f=2+1+1$ dynamical simulations 
recently carried out by the ETM Collaboration~\cite{Baron:2010bv}.
Our calculations were performed at one value of the 
lattice spacing ($a\approx 0.078 ~\fm$) and for fixed values of the 
pion and Kaon mass ($m_{\rm PS} \approx 390~\mev$ and 
$m_K \approx 580~\mev$, respectively). 

In evaluating these nucleon matrix elements, maximally 
twisted mass fermions are very helpful in two respects. 
The first is that the twisted mass fermion regularization provides a 
framework where it is possible to efficiently evaluate quark-disconnected diagrams. 
The second is that a consistent lattice framework can be set up where the matrix elements of interest are multiplicatively renormalizable  
and at the same time O($a$) improved.
As a result of these technical benefits we have been able to control the disconnected contributions and provide statistically
significant values for $\sigma_{\pi N}= m_l \la N|  \bar{u} u+  \bar{d} d |N \ra   $ and $\sigma_0 = m_l \la N|  \bar{u} u+ \bar{d} d  - 2 \bar{s}s|N \ra$.

In the case of the scalar charm content of the nucleon, our statistics was not sufficiently large to yield a signal above the statistical 
noise. We could thus only give the bound $|\la N|O_c|N \ra| \lesssim |\la N|O_s|N \ra|$.  

We remark that for phenomenological applications, the relevant quantities are 
actually the RGI quantities $m_s\la N|O_s|N\ra $ and $m_c \la N | O_c| N \ra$. 
Given our current statistical accuracy, it is therefore unclear at this moment whether the large 
Yukawa coupling of the charm quark can compensate the smallness of the matrix element.  

The most important achievement of this paper is the rather accurate evaluation of the ratio
$y_N=  2\la N| \bar{s} s |N \ra / \la N| \bar{u} u+ \bar{d} d |N \ra$  between the strange and the light quark content of the nucleon 
(see \eq{eq:y_para}), for which we find $y_N = 0.082(16)(2)$. 
The value we obtain is small compared to estimates from chiral perturbation theory  but is in line with recent lattice results obtained by other groups~\cite{Takeda:2010cw,Bali:2011rs,Horsley:2011aa,Durr:2011mp}.

Naturally the results presented in this paper need to be further scrutinized. 
In particular a careful study of the unwanted excited state contamination must be 
carried out to reduce the magnitude of the systematic errors associated to these 
effects. Finally data points at various lattice spacings and pion masses are necessary to be able to safely perform an extrapolation to the continuum limit and to the physical pion mass. 

\section*{Acknowledgments}
We thank G. Bali, K. Ottnad and C. Urbach for useful discussions. We thank all the members of ETMC for an enjoyable collaboration and for many fruitful discussions. G.H. acknowledges the support from the Spanish Ministry for Education and Science project FPA2009-09017, the Consolider-Ingenio 2010 Programme CPAN (CSD2007-00042), the Comunidad Aut\'onoma de Madrid (HEPHACOS P-ESP-00346 and HEPHACOS S2009/ESP-1473) and the European project STRONGnet (PITN-GA-2009-238353).  This work was performed using HPC resources provided by the JSC Forschungszentrum J\"ulich on the JuGene supercomputer and by GENCI (IDRIS-CINES) Grant 2010-052. It is supported in part by  the DFG Sonder\-for\-schungs\-be\-reich/Trans\-regio SFB/TR9.

\newpage
\appendix

\section{Scalar density renormalization with maximally twisted Wilson quarks}
\label{sec:MTWF}

In this Appendix we want to discuss the renormalization properties of scalar quark operators. We will separately discuss the unitary and the mixed action case in the setting offered by maximally twisted Wilson fermions~\cite{Frezzotti:2000nk}. 

We work here in the so-called physical quark basis and adopt
the notations of refs.~\cite{Frezzotti:2003ni,Frezzotti:2003xj,Frezzotti:2004wz}.
We recall however that, as it is customary (see e.g.\ ref.~\cite{Constantinou:2010gr}), 
the operator renormalization constants (RC's), being independent of
the twisting angle (in all mass-independent schemes),
are named after the form operators take in the twisted quark basis.

\subsection{Unitary degenerate doublet}
\label{sec:UD}

It has been proved in refs.~\cite{Frezzotti:2000nk,Frezzotti:2003ni} that at maximal twist\footnote{We recall that at maximal twist the flavour group of the lattice theory is the so-called $SU(2)_{\rm oblique}$ group with generators $\{Q_A^{1}, Q_A^{2}, Q_V^{3}\}$.} in the case of a degenerate $u$, $d$ doublet, 
with $\mu_u=-\mu_d\equiv \mu_l$, quark mass and scalar density renormalize according to the formulae 

\beqn
&&\mu_l^R=Z_P^{-1}\mu_l \, ,\label{MDD}\\
&&(\bar u u+\bar d d)^R=Z_P \Big{[}\bar u u +\bar d d - c_S(g_0^2,a^2\mu^2_l)\frac{\mu_l}{a^2}1\!\!1\Big{]}  \, ,\label{UDMT}
\eeqn
where \eq{UDMT} is written in the physical quark basis (with $r_u=-r_d$ as 
a consequence of the above chosen values of $\mu_u$ and $\mu_d$) and the last term in 
its r.h.s.\ represents the mixing of the quark scalar density operator with the 
identity. 

In this paper this term is of no importance because it will be automatically subtracted out in the computation of the nucleon matrix elements as described in the main text. For this reason, in order not to overload the forthcoming formulae, this mixing will not be indicated anymore.

\subsection{Unitary non-degenerate doublet}
\label{sec:UND}

For a pair of maximally twisted mass non-degenerate quarks, 
which for concreteness we name $s$ and $c$, 
one finds~\cite{Frezzotti:2003ni,Frezzotti:2004wz} the more complicated relations  

\beqn
&&\mu_c^R\equiv \bar m^R+\epsilon^R = Z_P^{-1}\bar m+Z_S^{-1}\epsilon\, ,\label{MNDDP}\\
&&\mu_s^R \equiv \bar m^R-\epsilon^R = Z_P^{-1}\bar m-Z_S^{-1}\epsilon\, ,\label{MNDDM}
\eeqn
\beqn
&&(\bar c c)^R=\frac{Z_P}{2} (\bar c c +\bar s s)+\frac{Z_S}{2} (\bar c c -\bar s s)\, ,\label{URTM}\\
&&(\bar s s)^R=\frac{Z_P}{2} (\bar u u +\bar s s)-\frac{Z_S}{2} (\bar u u -\bar s s)  \, ,\label{DRTM}
\eeqn
where the bare mass parameters $\bar m$ and $\epsilon$ coincide, respectively, with
the average mass and the mass difference of $c$ and $s$ quarks in free theory.

\subsection{OS valence fermions}
\label{sec:OS}

The mass RC of each OS valence fermion in the action is $Z_\mu=Z_P^{-1}$. This follows from the proof provided in  ref.~\cite{Frezzotti:2004wz} or from the extension of an old argument given in ref.~\cite{Maiani:1987by} that we reproduce in sect.~\ref{sec:APPB} for completeness. We thus get
\beq
\mu_{OS}^R=Z_P^{-1}\mu_{OS} \, .\label{OSM}
\eeq
We recall that $Z_P$ is an even function of the $r$-Wilson parameter.

In the philosophy of the ``mixed action'' approach proposed in ref.~\cite{Frezzotti:2004wz},  
a pair of mass degenerate OS fermions, denoted (in the so-called physical basis) as 
$s_+$ and $s_-$, with opposite values of the Wilson parameter ($r_{s_+} = -r_{s_-}=1$)
is introduced 
to represent the valence $s$ quark, with the understanding that no Wick contractions 
between the fermion $s_+$ and the fermion $s_-$ is allowed. This is done also in
sect.~\ref{eq:Nucleon_scalar}, with $\mu_s>0$ denoting the bare $s$ quark mass.

Consider a correlator where besides the strange quark scalar density only (renormalized) operators containing no strange quark are present. Then the insertion of the renormalized combination (we recall that the divergent mixing with the identity must be subtracted out)
\beq
(\bar s s)^R=\frac{Z_P}{2} [\bar s_+ s_+ + \bar s_- s_-] \, ,
\qquad r_{s_+} = -r_{s_-}=1 \,  .\label{SSPTM}
\eeq 
is finite, i.e.\ no new divergences are introduced. 
The reason for this fact can be traced back to the cancellation of chiral violating effects (coming from ``quark disconnected" - i.e.\ OZI~\cite{Okubo:1963fa,Zweig:1964jf,Iizuka:1966wu} violating - diagrams) between the two self-contractions (``loops") of the two valence quarks regularized with opposite values of $r$. 
Alternatively this result can be ascribed to the fact that, having the members of the $s_+$, $s_-$ pair opposite values of the $r$ parameters they look like a mass degenerate (valence) flavour doublet (in the "physical'' quark basis), e.g.\ just as the mass degenerate $u$ and $d$ pair discussed above. Naturally it remains the fact that the theory is not unitary since valence and sea quarks are regularized differently. This lack of unitarity leaves behind only O($a^2$) effects~\cite{Frezzotti:2004wz}.  

\section{Mass renormalization for OS valence fermions}
\label{sec:APPB}

For completeness in this appendix we want to explicitly prove the relation~(\ref{OSM})
along the lines of ref.~\cite{Maiani:1987by} in a setting where $N_v\geq 2$ valence OS fermions are introduced over an $SU(N_f=2)$ (or $SU(N_f=2+1+1)$) maximally twisted sea. The renormalization condition~(\ref{OSM}) is valid for anyone of the $N_v$ OS fermions, being $Z_P$ the renormalization constant of the non-singlet pseudo-scalar quark density.

The key observation to prove Eq.~(\ref{OSM}) is to recall that in order to be entitled to use the techniques that are usually employed to derive WTIs, it is necessary to have a fully local formulation of the theory. This means that for the purpose of dealing with a mixed action case, one has to keep in mind that for each OS valence quark in the action a corresponding ghost with equal mass (and opposite statistics) has to be introduced. Without loss of generality, for the purpose of proving Eq.~(\ref{OSM}), we can assume that all valence quarks (and ghosts) have the same bare (twisted) mass, $\mu_{OS}$. The generalization to the non-degenerate valence quarks is straightforward. 

\subsection*{WTIs for OS fermions}
\label{sec:SAPPB}

Let $A_\mu^a$, $a=1,2, \ldots, N_v^2-1$ be the (non-singlet) axial vector current constructed in terms of only valence fermions (and no ghosts) and let us assume that we are exactly at maximal twist. 
For convenience we shall work in the twisted fermion basis where 
the valence quark mass term has the expression 
\beq
L^{OS}_{mass}=a^4\sum_{x}\mu_{OS}\Big{[}\sum_f (\bar \chi_f(x) i\gamma_5 \chi_f(x) + {\mbox{ghosts}}) \Big{]}\, .
\label{MASTE}
\eeq
The argument can be split into four parts

I - The (bare/lattice) axial WTI between on-shell states reads~\cite{Bochicchio:1985xa}
\beq
\langle \alpha|\nabla_\mu A_\mu^a(0) |\beta\rangle=2\mu_{OS} \langle \alpha|S^a(0)|\beta\rangle -\langle \alpha|X^a(0)|\beta\rangle \, ,
\label{WTIOS}
\eeq
where 
\beq
S^a=\bar \chi \tau^a\chi 
\label{MASTEA}
\eeq
and $X^a$ is the chiral variation of the OS Wilson term, the explicit expression of which we do not need in this discussion. The only thing we need to know about $X^a$ is its mixing pattern (dictated by dimensional argument and the symmetry ${\cal{P}}_\chi \times (\mu_{OS}\to -\mu_{OS}$), where ${\cal{P}}_\chi$ is the formal parity acting on the twisted fields, see ref.~\cite{Frezzotti:2000nk}) which reads 
\beq
X^a(x)=(Z_A-1)\nabla_\mu A_\mu^a(x) +2 \eta_X\mu_{OS} S^a(x)+\bar X^a(x)\, , \label{CHI}
\eeq
with $\eta_X$ a (finite) function of $g^2_0$. Inserting Eq.~(\ref{CHI}) in~(\ref{WTIOS}), one gets 
\beq
\langle \alpha|Z_A\nabla_\mu A_\mu^a(0)|\beta\rangle=2(1-\eta_X)\mu_{OS} \langle \alpha|S^a(0)|\beta\rangle -\langle \alpha|\bar X^a(0)|\beta\rangle \, ,\label{WTIBB}
\eeq
Since between on-shell states $\bar X^a$ can only give rise to O($a$) terms, the renormalized (continuum looking) WTI
\beq
\langle \alpha|Z_A\nabla_\mu A_\mu^a(0)|\beta\rangle=2\mu^R_{OS} \langle \alpha|S_R^a(0)|\beta\rangle +\mbox{O}(a) \, .\label{WTIBC}
\eeq
is immediately obtained by setting
\beqn
&&S^{aR}=Z_S S^a \, ,\label{SRIN}\\
&&\mu_{OS}^R= (1-\eta_X)Z_S^ {-1}\mu_{OS} \equiv Z_\mu \mu_{OS} \, .\label{MRIN}
\eeqn 
Our aim is to prove the relation 
\beq
Z_P^ {-1}=(1-\eta_X)Z_S^ {-1}  \label{ZPS}
\eeq
from which the formula
\beq
Z_P^ {-1}=Z_\mu  \label{ZMU}
\eeq
follows.

II - To this end we need to extend the previous equations to the case where the divergence of the axial current is inserted together with the singlet pseudo-scalar density 
\beq
\tilde P_0=\sum_f(\bar \chi_fi\gamma_5\chi_f +{\mbox{ghosts}})\label{PTIL}
\eeq
which results from considering contributions coming from valence quarks as well as the associated ghosts. We note that as the axial current we are considering is only made up of valence quarks, it cannot rotate the ghost fields. We thus get for the WTI where the operator $\tilde P_0$ is inserted
\beqn
&&\langle \alpha|\nabla_\mu A_\mu^a(x) \tilde P^0(y)|\beta\rangle=2\langle \alpha|S^a(x)|\beta\rangle \delta (x-y)+\nn\\
&&+2\mu_ {OS}\langle \alpha|S^a(x) \tilde P^0(y)| \beta\rangle -\langle \alpha|X^a(x)\tilde P^0(y)| \beta\rangle \, ,\label{WTIB}
\eeqn
with external states such that one does not get identically vanishing matrix elements. Introducing the decomposition~(\ref{CHI}), we first rewrite the previous equation in the form 
\beqn
&&\langle \alpha|Z_A\nabla_\mu A_\mu^a(x) \tilde P_0(y)| \beta\rangle=2\langle \alpha|S^a(x)|\beta\rangle \delta (x-y)+\nn\\
&&+2(1-\eta_X)\mu_{OS}\langle \alpha|S^a(x) \tilde P_0(y)| \beta\rangle -\langle \alpha|\bar X^a(x) \tilde P_0(y)|\beta\rangle \, .\label{WTIBF}
\eeqn
We remark that, when $\bar X^a$ is inserted with a local operator, it gives rise 
to localized terms plus genuinely O($a$) contributions. Simple symmetry considerations and the fact that in $X^a$ only valence  fermions (and not ghosts) appear, imply
\beq
\langle \alpha|\bar X^a(x) \tilde P_0(y)| \beta\rangle=2c_X\langle \alpha|S^a(x)| \beta\rangle \delta (x-y)+{\mbox{O}}(a)\, ,\label{XBAR}
\eeq 
where $c_X$ is again a finite function of $g^2_0$. Substituting into Eq.~(\ref{WTIBF}) and neglecting irrelevant (for the present argument) O($a$) terms gives 
\beqn
&&\langle \alpha|Z_A\nabla_\mu A_\mu^a(x) \tilde P^R_0(y)| \beta\rangle=2(1-c_X)Z_{\tilde P_0} \langle \alpha|S^a(x)|\beta\rangle \delta (x-y)+\nn\\
&&+2\mu^R_{OS}\langle \alpha|S_R^a(x) \tilde P^R_0(y)| \beta\rangle  \, ,\label{WTIBCONT}
\eeqn
where we have multiplied both members by $Z_{\tilde P_0}$ and used Eqs.~(\ref{SRIN}) and~(\ref{MRIN}). It must be stressed that consistency with continuum WTIs (universality) at vanishing $\mu_{OS}$~\cite{Bochicchio:1985xa,Testa:1998ez} requires the identification 
\beq
Z_{\tilde P^0}(1-c_X)=Z_S\, .\label{XBARR}
\eeq

III - The third step of this analysis is inspired by the discussion carried out at the end of sect.~2 of ref.~\cite{Maiani:1987by}. One notices that by summing over $x$ in both members of Eq.~(\ref{WTIBCONT}), one gets at $\mu_{OS}\neq 0$ the identity 
\beq
0=2\langle \alpha|S^{aR}(y)| \beta\rangle +2\mu_{OS}^R \langle \alpha|\sum_x S^{aR}(x) \tilde P^R_0(y)| \beta\rangle \, .\label{WTIB2}
\eeq
as the integral of a divergence vanishes. Once also summed over $y$, Eq.~(\ref{WTIB2}) can be usefully compared to the formula one gets from the obvious identity
\beq
0=\frac{\partial}{\partial \mu_{OS}}\langle \alpha|\sum_x Z_A \nabla_\mu A_\mu^{a}(x) |\beta\rangle=
\frac{\partial}{\partial \mu_{OS}}\langle \alpha|\sum_x  2\mu_{OS}^R S^{aR}(x) | \beta\rangle\, ,\label{WTIB4}
\eeq
in which the second equality follows from Eq.~(\ref{WTIBC}). Indeed by explicitly performing the derivative with respect to $\mu_{OS}$, one finds 
\beq
0=2(1-\eta_X)Z_S^{-1}\langle \alpha|\sum_x S^{aR}(x)| \beta\rangle +2\mu_{OS}^R \langle \alpha|\sum_x S^{aR}(x)\sum_y \tilde P_0(y)| \beta\rangle \, .\label{WTIB5}
\eeq
Multiplying this equation by $Z_{\tilde P_0}$ and comparing with~(\ref{WTIB2}) summed over $y$, one gets 
\beq
Z_{\tilde P_0}(1-\eta_X)= Z_S\, .\label{ZPSR}
\eeq
We remark that Eq.~(\ref{ZPSR}) taken together with Eq.~(\ref{XBARR}) entails the somewhat surprising equality $\eta_X=c_X$.

IV - As the final step of this long argument we want to now show that
\beq
Z_{\tilde P_0}=Z_{P} \label{ZPPT}
\eeq
which would finally prove Eq.~(\ref{ZPS}). The reason for the validity of Eq.~(\ref{ZPPT}) is that, when one considers the diagrams contributing to $Z_{\tilde P_0}$, one realizes that the OZI-violating diagrams where a valence quark is self-contracted (closed into a ``loop'') is exactly cancelled by the contribution where the corresponding ghost, present in $\tilde P_0$, is closed into a ``loop''. One is thus only left with diagrams in which neither valence nor ghost self-contractions appear, hence exactly with the diagrams that  contribute to the non-singlet pseudo-scalar density RC, $Z_P$, where such self-contractions are forbidden by flavour conservation.

\bibliography{strange_content}

\end{document}